# Safety, Security, and Privacy Threats Posed by Accelerating Trends in the Internet of Things

Kevin Fu, Tadayoshi Kohno, Daniel Lopresti, Elizabeth Mynatt, Klara Nahrstedt, Shwetak Patel,
Debra Richardson, Ben Zorn

**Abstract:** The Internet of Things (IoT) is already transforming industries, cities, and homes. The economic value of this transformation across all industries is estimated to be trillions of dollars and the societal impact on energy efficiency, health, and productivity are enormous. Alongside potential benefits of interconnected smart devices comes increased risk and potential for abuse when embedding sensing and intelligence into every device. One of the core problems with the increasing number of IoT devices is the increased complexity that is required to operate them safely and securely. This increased complexity creates new safety, security, privacy, and usability challenges far beyond the difficult challenges individuals face just securing a single device. We highlight some of the negative trends that smart devices and collections of devices cause and we argue that issues related to security, physical safety, privacy, and usability are tightly interconnected and solutions that address all four simultaneously are needed. Tight safety and security standards for individual devices based on existing technology are needed. Likewise research that determines the best way for individuals to confidently manage collections of devices must guide the future deployments of such systems.

## Introduction

Increasingly we live in a world of connected smart devices. This "Internet of Things" (IoT) combines devices with sensor capabilities and connectivity to the cloud and allows them to leverage artificial intelligence, machine learning, and big data analytics, sometimes dramatically increasing their capabilities. Everyday users have progressed from having a single home computer to a variety of devices that are each individually managed, which can be difficult. For example, due to consumers failing to change the default password, many baby monitors allow arbitrary strangers on the web to view unsuspecting people's homes.

But the proliferation, capabilities, and interconnectedness of smart devices present dramatic new opportunities and challenges that require new research and industry approaches to make such systems safe, secure, effective, and usable. The problem is so acute that the FBI recently issued a public service announcement suggesting consumers should "Isolate IoT devices on their own protected networks" and "...be aware of the capabilities of the devices…" which are expectations highly unlikely to be followed in practice. In this paper, we argue that collections of smart devices present new challenges that require a greater understanding of how people can effectively use such systems and a deeper investment in policies and tools that give users confidence in them. In particular, issues related to security, physical safety, privacy, and usability are tightly interconnected and solutions that address all four simultaneously are needed.

There have been numerous estimates of the impact of the Internet of Things on the economy, with estimates that the number of deployed devices will be 50 billion by the year 2020 and that the total economic impact may be up to 10 trillion dollars by 2025. We already live in a world of interconnected devices, with numerous companies offering smart devices such as smart thermostats, smart doorbells, etc. In the so-called Industrial Internet of Things and Smart Cities initiatives, factories and cities will become infiltrated with interconnected smart devices, with large projected improvements in efficiency and reliability. For example, hospitals have benefitted from a proliferation of interconnected sensor devices, resulting in improved health outcomes and lower costs.

Unfortunately, as the number and connectivity of such devices increases, the challenge of managing these collections of devices becomes exponentially more difficult. If managing a single home computer is difficult for a non-technical person, imagine what is needed to understand and correctly manage a network of many interacting devices? For example, consider a hypothetical scenario where an Apple iPhone, a Ring doorbell, an Amazon Echo and an Xbox work together. The iPhone is used the configure them and the Echo is used to to implement voice commands so that, for example, a user could tell Echo to show the video feed from the Ring on the TV using the Xbox. Another example could focus on energy usage and home monitoring. Smart water and electricity meters could coordinate to monitor and adjust water and power while determining what patterns of home activity correlate to high usage.

Making a single device secure and safe is already a difficult problem. Safety issues, in particular, are increasingly important for IoT systems as they are used to physically control electrical devices like light bulbs and heating systems in both homes and in businesses. The safety problems discovered with the Samsung Galaxy Note smartphones catching fire illustrate the challenges making devices safe even without an attacker trying to cause harm. Making them safe in the presence of an attacker is even more difficult and requires rethinking how such devices are designed and tested for safety. The consequences of having many insecure individual devices attached to the internet was highlighted recently when the Marai malware was used to create a 380,000 IoT-based botnet used in a massive distributed denial of service (DDOS) attack. Only a month later, a major cyberattack harnessed tens of millions of machines, including a large number of IoT devices, aimed at the Internet's domain name server (DNS) infrastructure, disrupting a number of major service providers including Twitter, Netflix, Spotify, Airbnb, Reddit, Etsy, SoundCloud and The New York Times. As more insecure nodes are attached, the leverage an attacker gets in using them increases.

Beyond the existing challenges of securing individual devices, we need to simplify how people interact with a collection of devices so that they don't have to think about each device and how they might interact. For example, with an iPhone, Echo, Ring, and Xbox, what information is being shared between the devices and what are the privacy policies in place regarding what information from a private home can be sent to the different companies and how can this information be used? Beyond privacy, what security vulnerabilities does this particular collection of devices create and what entity is responsible for informing owners that such vulnerabilities exist? In much the same way that operating systems have evolved to allow individual users to configure and manage them, new technology is needed for users to more easily understand, configure, and manage their collections of devices.

In this paper, we consider two scenarios where collections of devices create opportunities and challenges: interconnected devices in a smart home and device collections in hospitals. By looking at both a consumer-oriented scenario and safety-critical commercial applications, we can observe similarities and differences in the requirements for such systems.

**The Internet cartoon Joy of Tech's interpretation of the future of IoT**

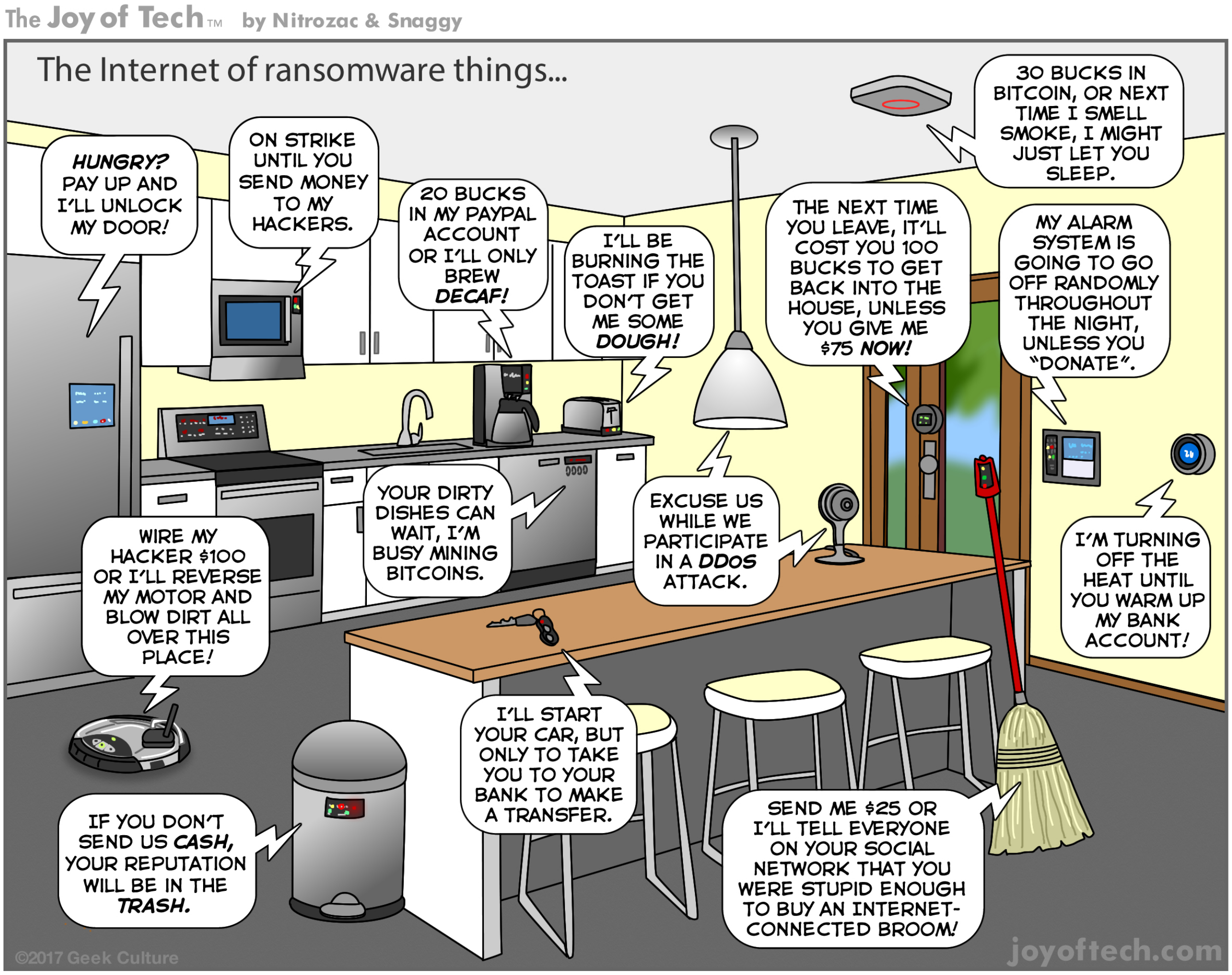

## Smart Devices in Homes

Despite the availability of many connected solutions for the home, the rapid growth of this space has outpaced security and privacy research, regulatory guidelines, discussions on longevity and safety, and a general understanding of how such systems reflect human understanding and mental models. However, the emergence of scalable smart home systems has the potential to directly impact our daily lives. Thus, we present a set of opportunities and challenges for computing research for smart home technology.

With more and more connected appliances appearing on the market—such as Jarden's Mr. Coffee™ and Crock-Pot™—new physical safety hazards emerge due to the ability for software to control these high-powered loads. Recent work has shown the safety hazards of simple WiFi-enabled appliance modules and light bulbs. Analogous to mandated safety measures such as electrical circuit breakers, GFCI switches, and fire-rated walls that protect consumers from faults in home infrastructure, smart home technologies need a similar layer of protection. Just as National Electric

Codes (NEC) and National Electrical Manufacturers Association (NEMA) exist to provide safety guidelines, similar safety enforcement processes need to evolve for IoT appliances in the home. Building codes will also need to evolve to support emerging smart home technologies. Addressing safety hazards for home IoT devices will require a coordinated effort between the computing community and the Department of Housing, Federal Communications Commision (FCC), Underwriters Laboratories (UL), and National Institutes of Standards and Technology (NIST).

Smart home technologies, and the IoT in general, pose a new challenge in abandonment by manufacturers, especially IoT startups that may introduce a product in the market and quickly go out of business or completely abandon support. These so called “zombie” devices remain on a home network without future support for security and safety patches. These risks are problematic for technologies that are integrated into the home’s infrastructure or appliances that may reside in the home for many years, creating both a policy and a technology challenge. There is a need for approaches to effectively detect these abandoned systems and monitor the interaction of these devices with other platforms. The other extreme would be to require manufacturers to remotely disable legacy devices when support ceases.

## Smart Devices in Hospitals

Hospitals – and healthcare in general – benefit greatly from computation. Computation can enable more accurate, more informed patient care in the form of electronic medical records. Computation enables increased efficiency within hospitals, allowing a single nursing station to wirelessly monitor many patients at once. For example, a nursing station could remotely – and wirelessly – monitor the drug pumps dispensing drugs to all the patients within their care. Computation even occurs inside patients’ bodies in the form of wireless implantable medical devices, like pacemakers and implantable cardiac defibrillators.

Unfortunately, it has long been known that with the increased benefits of computation in hospitals also comes the potential for patient harm if there are defects in the systems’ software. A canonical example is that of the Therac-25, a radiation therapy device from the 1980s that was found to have a software defect that could cause patients to receive approximately 100 times the radiation therapy that they were supposed to receive. This software defect, human factors, and project mismanagement resulted in harm to patients, and at least several deaths. These harms were caused by accident. In the cyber security arena, we must ask: what might an intelligent, creative adversary be able to accomplish, and how can we provide resiliency against such an adversary. That adversary can clearly cause at least as much harm as might occur by accident, and likely more, because that adversary can force the systems into their worst-possible configurations. Moreover, due to the increased pervasiveness of computation within the healthcare environment, the potential attack surface to cyber adversaries is even greater today than it was the 1980s.

A comprehensive approach to cyber security in hospitals must consider each of the computational devices within the hospitals, as well as what those devices depend on. For example, cyber attacks against the hospital’s power infrastructure could significantly impact patient care. Cyber attacks against the hospital’s water supply could also significantly impact patient care. There have been cases where hospital servers have been shut down by ransomware, thereby requiring healthcare providers to revert to paper-based records – something that many younger hospital staff might not be trained to work with. Building on the ransomware scenario, imagine the impact of even more malicious malware, such as malware that intentionally modifies patient electronic prescriptions or dosages records, to possibly dangerous drugs or drug levels. One can similarly imagine the potential impact of compromising hospital devices that directly impact patient care, ranging from computerized radiation therapy devices to the devices that doctors use to wirelessly change the settings on implantable medical devices, like pacemakers and implantable drug pumps.

We stress that cyber security is about risk management, and that the set of harms that might be possible is often greater than the set of harms that are likely to occur in practice. Hospitals – and healthcare in general – need to be vigilant in assessing the spectrum of potential harms so that they are not surprised by unexpected impacts, and then realistic about assessing the actual risk of these harms. Security best practices should be used whenever possible. For example, devices should not use default passwords. And, when possible, if a device is known to have a cyber vulnerability, then that device should receive a software update.

## Smart Health in the Home

The previous two scenarios combine in interesting ways when one considers the increasing use of healthcare technologies in the home. Whether motivated by sustaining older adults wishing to "age in place," the increasing use of wearable sensors (now often worn before and after surgical treatment), or the increasing interest in accountable care and the need to monitor patients "in the wild" to help ensure treatment success, digital technologies are seeping out of traditional healthcare environments and finding their way to typical homes. In this perfect storm, we now have the safety and security vulnerabilities combined as two systems (home and healthcare) attempt to reside in the same physical setting and likely on the same wireless network. The home becomes a backdoor vulnerability to the hospital and visa versa.

What is at stake, beyond security, is the desired reliance on data generated in the home to inform healthcare decision making. This data could be paramount in helping older adults avoid the costs of institutional care, in helping patients undergoing treatment to stay out of emergency rooms when not needed, and getting to them when critical, and helping patients whose illness includes environmental triggers (e.g. asthma) manage their treatment and behavior on a day to day basis.

## Implications of the Scenarios

### Security and Physical Safety

The most important requirement for collections of devices is that they guarantee physical safety and personal security. While there has been a great deal of research and commercial investment in preventing cyberattacks, protecting collections of devices presents new challenges that have not been addressed. In particular, the ability of smart devices to control physical aspects of the environment (such as the house temperature or whether a door is locked) creates potential attacks on an individual's physical safety that requires even higher levels of assurance than existing cyberattack countermeasures. The distributed and interconnected nature of multiple systems present in device collections also requires rethinking of the basic concept of security and system management. Without taking a multi-system view, security techniques will be unable to anticipate and counter vulnerabilities that arise from incorrect configurations or attacks that exploit vulnerabilities in the way that devices interact with each other and with computing in the cloud.

Because interacting devices have been present in hospitals for some time, and because hospitals are subject to regulatory frameworks that require higher levels of compliance, the hospital scenario for managing collections of smart devices is better understood. Insights based on this experience include: (a) the life-cycle of the device, including how software is upgraded, must be taken into consideration, (b) physical accessibility of devices, including the ability for an intruder to access interfaces such as USB ports or Wifi networks, must be carefully controlled, and (c) the regulatory framework around privacy makes reasoning about where data is collected, how it is shared, and where it is stored very challenging.

Contrasting the two scenarios of devices in the home versus devices in a hospital, we draw several conclusions. First, different degrees of security vetting and analysis are required for each scenario. There are already regulatory constraints on medical devices but the exploding complexity and increasing potential vulnerabilities require thoughtful revisiting of what level of certification is required to provide appropriate levels of security and safety assurance for such applications. The recent news of security vulnerabilities in St. Jude pacemaker devices highlights the challenges in determining the right level of cybersecurity assurance needed for individual devices and also the overall collection of devices. Likewise, hospitals would be more attractive targets for coordinated attacks akin to current “ransomware” attacks currently being conducted on hospital electronic health record (EHR) systems. Second, while hospitals employ IT professionals to manage their collections of devices, consumers have no such support but are subjected to similar challenging system complexity. The recent report from the Commission on Enhancing National Cybersecurity highlights similar risks to small businesses that cannot afford an IT staff. Any improvements in allowing individuals to understand and manage such a collection of devices will benefit both scenarios but the consumer scenario requires rethinking how such systems can be explained in terms accessible to everyday users.

## Privacy

Privacy is challenging to understand and guarantee in a world where more and more smart devices collect data, share it, and monetize it. The model that software is monetized by advertising is being applied at the device level. Many free smartphone apps already collect data at the user’s expense and sell it in ways that are not obvious or explicit to the consumer. Algorithmic techniques such as differential privacy provide theoretical assurances to limiting the potential impact of data sharing, but such techniques are rarely used in practice and as a result the privacy implications of increasingly intrusive smart devices and sensors are unknown. The complexity of understanding the privacy policy of a single application, like Facebook, can overwhelm individual users and the burden of understanding such policies for every device and application being used requires attention and complexity beyond most people.

Consider, then, the challenge of understanding not just one device but many that interact in complex ways. Without new mechanisms for explaining what information is being collected and shared, not by each individual device, but in aggregate, users will be unable to understand what the privacy implications of their choices are. Consider, for example, buying a smart fork (a real device). How does a consumer know what information the fork is collecting (beyond counting the individual fork lifts, for example)? What if the consumer then buys a smart plate? Can the fork and plate exchange information? And if so, what can be inferred from the combination of the information that can’t be determined from either data source? Consider for example an Internet TV service and a smart thermostat. The use of smartphones to control these devices creates data to identify individuals in the home. The thermostat can then pinpoint who is where in the home and when. A few IoT devices in the home can lay out a pretty detailed map and timeline of home activities.

In the hospital setting regulatory compliance with HIPPA and other regulations determines what is legal regarding data collection and sharing. The complexities of understanding whether a particular device configuration is compliant relies on the wisdom and understanding of IT professionals. As the complexity of data being collected increases and the ways it is used become more diverse, really understanding the privacy implications of a particular configuration is likely to challenge even the best-informed IT professionals.

Beyond understanding privacy implications of connected devices acting as they are intended, the implications of data breaches on privacy due to security vulnerabilities increases the complexity and risk in providing adequate privacy guarantees. Fortunately, advances in storing and operating on encrypted data will likely provide technical solutions to some of the challenges of preventing data breaches. Nevertheless, the presence of malicious state-sponsored actors attacking the privacy of high-profile individuals greatly increases the level of protection needed to provide overall confidence in such systems. Ultimately, social engineering attacks and attacks based on inadequate human understanding of these systems remains perhaps the greatest challenge to overcome.

### Usability and the User Experience

We have already made the case that the ability for professionals or consumers to understand and manage complex systems creates significant vulnerabilities to security, safety, and privacy. To attack this problem there are two approaches: either simplify the systems sufficiently that they can then be understood, or build better conceptual models for users and tools to reduce the burden. Due to the widespread use of open-source software including Linux in creating many smart devices, the configuration of many smart devices is arcane and assumes significant expertise to understand and manage. Simplifications can be made by reducing the number of choices and exposing the configuration as a "wizard" but there are limits to what can be eliminated. Another simplification is to explicitly disallow devices from interacting with each other. While this scheme reduces the management burden of the user, it also significantly reduces the potential value of the system. For example, a device that determines that there is no one present in a house might want to communicate with the device controlling a garage door to close it, but their interaction would be prevented.

As an alternative, new approaches to helping individuals see the bigger picture of their entire device collection is possible. In particular, a "device dashboard" might present a view of all the devices, how each is configured, and how they relate. Such a view can extend familiar concepts that users have in managing individual computers, such as security and privacy settings, to understanding their entire network. With such an aggregate view, tools that help users track the configuration, such as individual software updates, and guarantee the current configuration is secure can be developed and marketed.

Understanding how people think about technology, their willingness to adopt it, and their challenges in maintaining it needs to be a critical part of smart device research and policy going forward. No level of software security is sufficient if the person configuring the system fails to provide adequate passwords or understand that the system is misconfigured. Historically the human dimension of design could be offloaded to expert IT professionals but increasingly these hard usability problems need to be handled directly by consumers.

## Recommendations

Based on this discussion, we recommend the following approach to expanding the research agenda and policy agenda based on advances in the Internet of Things and adhoc collections of smart devices.

**Broad conclusions**

- **Problems of security, privacy and usability cannot be considered separately** - they need to be considered together and federal investments should prioritize solutions that focus on augmenting a person's ability to understand and manage complex systems.
- The potential for risks to physical safety requires that minimum levels of cybersecurity assurance be defined and required for widespread device deployment.
- Milestones must be established for determining the level of analysis and testing required for smart device products (akin to targeted EPA emission requirements). Specifically improve:
    - The transparency of the software the devices are running for inspection and analysis
    - The level of testing and analysis required for certification
    - The level of hardening of the critical components (crypto, secure communication, secure update channels)

**Secure and manage individual devices**
Existing efforts such as the Cybersecurity Assurance Program and the Report of the Commission on Enhancing National Cybersecurity provide guidelines and requirements to help ensure that individual devices are sufficiently secured. Beyond the current investments we recommend:

- Revising safety requirements for internet-connected electrical devices with an emphasis on adversarial thinking, in order to limit the damage that a remote attacker with harmful intent is able to do.
- Increasing the emphasis on building software and hardware based on verified components. Program verification technology is advancing rapidly and increasingly complex subsystems, such as cryptographic implementations should be developed using state of the art verification tools.
- Increasing requirements for program analysis and testing tools to certify software deployments in smart devices, with different levels of analysis required depending on the degree to which physical safety might be threatened by the device.
- Improving software update requirements for devices that are deployed to allow software to be patched as new vulnerabilities are discovered.
- Updating mechanisms that are resistant to exploitation using state-of-the-art encryption.
- Creating cradle-to-grave requirements that specify what happens when devices are no longer being updated, for example, because to the company producing them went out of business.
- Supporting research to help users correctly maintain their devices and software.

**Managing collections of devices**
Very little has been specified regarding managing collections of devices despite the fact that they are increasingly present. As a starting point, we recommend the creation of:

- Explicit software that considers all the devices in a collection and presents an overview of them to a user (device dashboard).
- Management tools that allow the user to understand and change the configuration so that it remains secure over time.
- Simplifications in the complexity of configuration management that prevent users from common errors that create security or privacy errors.
- A user experience that leverages concepts that users are already familiar with in managing individual devices.

## Summary

Technology is rapidly evolving and having a greater impact on society than it has ever had with sensing and intelligence starting to be embedded in every device. The advances bring significant benefits to people, companies, and organizations, but until the technology is better understood, there are also associated risks. We have outlined some of the implications of these changes through a discussion of use-case scenarios and the dimensions of safety, security, and privacy. We believe that changes are happening with such speed and the level of risk and uncertainty is sufficiently high that investment in research that helps mitigate potential problems should be prioritized. The potential benefit to human lives, our national interests, and the economy is sufficient to warrant substantial research investments in making the technology as beneficial as possible.

For citation use: Fu K., Kohno T., Lopresti D., Mynatt E., Nahrstedt K., Patel S., Richardson D., & Zorn B., (2017). Safety, Security, and Privacy Threats Posed by Accelerating Trends in the Internet of Things. http://cra.org/ccc/resources/ccc-led-whitepapers/

*This material is based upon work supported by the National Science Foundation under Grant No. 1136993. Any opinions, findings, and conclusions or recommendations expressed in this material are those of the authors and do not necessarily reflect the views of the National Science Foundation.*